\RequirePackage{fix-cm}
\documentclass[smallextended]{svjour3} 
\usepackage{tabularx}
    \newcolumntype{L}{>{\raggedright\arraybackslash}X}
\usepackage{graphicx}
\usepackage{placeins}
\smartqed  
\usepackage[dvipsnames]{xcolor}
\usepackage{url}

\title{A Survey on Applications of Digital Human Avatars toward Virtual Co-presence}
\author{Matthew Korban  \and
        Xin Li
}

\institute{Department of Electrical and Computer Engineering, Louisiana State University, Baton Rouge 70803, U.S.A.\\
\email{acw6ze@virginia.edu (Matthew Korban), xinli@lsu.edu (Xin Li)}
}

\date{Latest Revision: May 2021}

\begin{document}

\maketitle

\begin{abstract}
This paper investigates different approaches to build and use digital human avatars toward interactive Virtual Co-presence (VCP) environments. 
We evaluate the evolution of technologies for creating VCP environments and how the advancement in Artificial Intelligence (AI) and Computer Graphics affect the quality of VCP environments.
We categorize different methods in the literature based on their applications and methodology and compare various groups and strategies based on their applications, contributions, and limitations. We also have a brief discussion about the approaches that other forms of human representation, rather than digital human avatars, have been utilized in VCP environments.
Our goal is to fill the gap in the research domain where there is a lack of literature review investigating different approaches for creating avatar-based VCP environments. 
We hope this study will be useful for future research involving human representation in VCP or Virtual Reality (VR) environments.
To the best of our knowledge, it is the first survey research that investigates avatar-based VCP environments. Specifically, the categorization methodology suggested in this paper for avatar-based methods is new.

\end{abstract}

\section{Introduction}

Remote communication has a crucial role in modern societies. 
An essential aspect of effective remote communication is co-presence, where multiple participants can see and interact with each other in a shared virtual environment so-called \emph{Virtual co-presence (VCP)}. The VCP is a sense of being with others in a virtual world that people are psychologically connected, and are available and accessible to others \cite{bulu2012place}. The VCP environments can be more engaging than text or voice-based chat in communications, collaborations, and training. 
Recently, such a virtual co-presence has been studied in remote education~\cite{Mustufa12VRForDistanceEdu}, training simulations \cite{hooper2019virtual,schmidt2019heidelberg}, therapy treatments~\cite{WiederholdVRET05Book} and social interaction venues \cite{hudson2019or}. 

A VCP environment can be created using a Virtual Reality (VR) platform. A well-known approach to represent the human character in VR environments is using ``Avatar'' models. An avatar is a digital form of human character that can be represented as 2D or 3D. 3D avatars have some advantages over 2D avatars, such as being more human-like, having realistic motions, and offering an immersive experience in VR and VCP environments. 

In this paper, we investigate different technologies designed toward creating VCP environments based on digital human avatar models. We evaluate how the evolution of technologies in Artificial Intelligence (AI) and Computer Graphics affect the human representation quality in VCP environments.
We mainly study two types of researches: (1) works that create or use digital avatars models in VCP environments (2) works that build the avatar models or the pipelines that are highly beneficial for developing VCP environments.
For both types of research, we will explain and compare their advantages, limitations, and also applications. 
We also will have a short discussion about the methods that used other forms of human character representation rather than digital human avatar to create VCP environments. 

\begin{figure*}[h!tbp]
	\centering
	\includegraphics[width=1.0\textwidth]{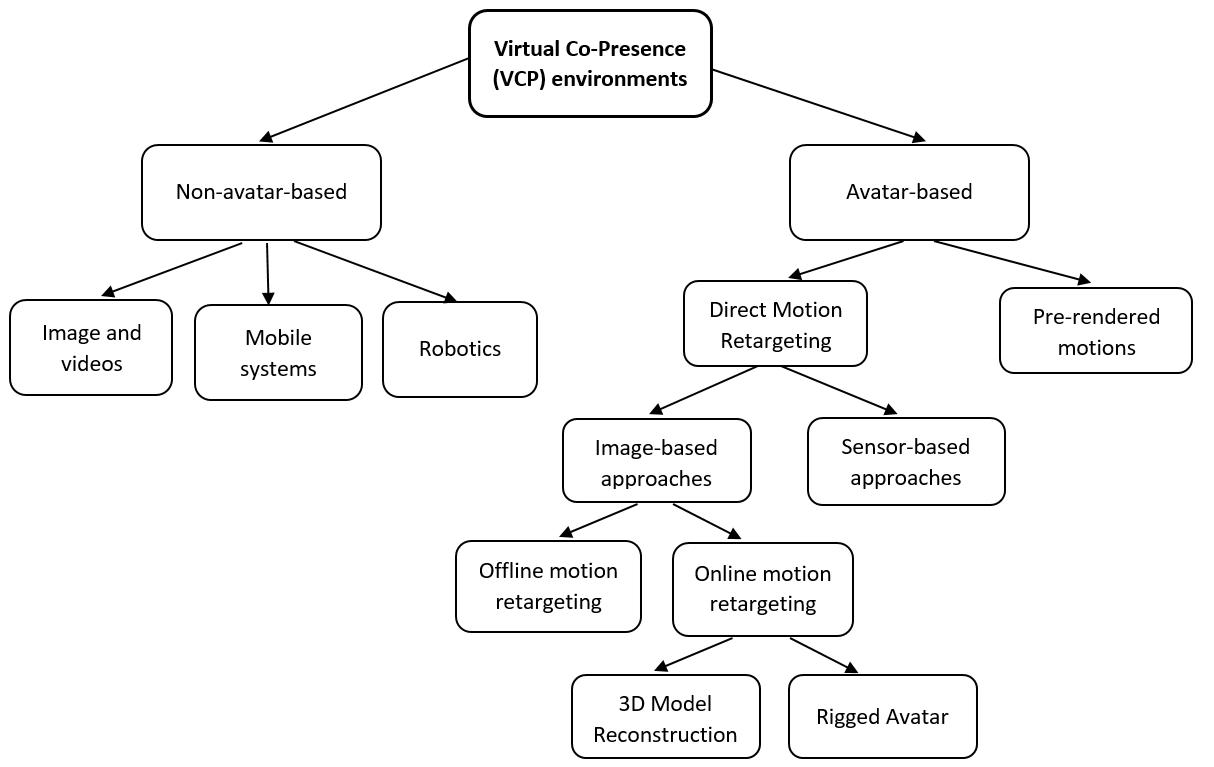}
	\caption{Our categorization methodology of different methods in the literature toward creating Virtual co-presence (VCP) environments. ~\label{Fig:survey}}
\end{figure*}

Fig. \ref{Fig:survey} shows our methodology of categorizing of the works carried out in the literature to construct VCP environments. As can be seen in the figure, we classify non-avatar based methods into three categories of \emph{Robotics}, \emph{Mobile systems}, and \emph{Images and videos}. We also compare non-avatar based researches based on their applications, advantages and disadvantages. 

On the other hand, we classify the avatar-based approaches into two main types of \emph{Direct-motion retargeting} and \emph{Pre-rendered motions}. In the methods that used Direct-motion retargeting, the motion is transferred directly from users to digital avatars. In contrast, in the approaches that exploited Pre-rendered motions, the motions are pre-defined by developers.
Direct-motion retargeting methods provide authentic motions resembling users' actions. Although some of these strategies are not used directly in VCP environments, their algorithm pipelines or avatar models are highly beneficial in creating prospective VCP environments based on digital human avatars.  

Next, we further divide the Direct-motion retargeting category into two classes of \emph{Image-based} and \emph{Sensor-based} approaches. In the image-based strategies, the input is commonly RGB images, video, or depth images 
On the other hand, in the sensor-based approaches, the input is obtained from wearable sensor devices, tracking sensors, and tracking markers.
Both Pre-rendered motion and Sensor-based categories mostly include similar application-based approaches. As a result, we compare these two categories based on their applications, benefits, and drawbacks.

The Image-based category consists of two groups of \emph{Offline motion retargeting} and \emph{Online motion retargeting}. Offline motion retargeting approaches often offer more accurate results than online methods by reconstructing high-quality body shape, pose, and textures. Online motion retargeting is, however, faster and more suitable for a real-time VCP or VR experiments. Furthermore, the online motion retargeting approaches consist of two types of (1) \emph{3D model reconstruction}, where they reconstruct the human character model or scene, and (2) the methods that used pre-designed \emph{Rigged avatar} models. 
In this paper, we compare different Image-based strategies based on their applications, advantages, and disadvantages in VCP environments. Since the image-based methods commonly made technical contributions, we also will illustrate the gradual evolution of technology in each sub-category. This illustration explains how new approaches resolve previous issues.  

There have been a few research reviewing different approaches to build VCP environments. Krist et al. \cite{kristoffersson2013review} are the only researchers that conducted a comprehensive study on VCP environments. However, they discussed the strategies that utilized non-avatar/robotics methods. On the other hand, in our research, we focus on using the digital human avatar, which is the state-of-the-art form of human representation in VCP environments. 
Moreover, there has also been a lack of thorough research on human avatars in general. Hudson et al. \cite{hudson2016avatar} conducted a short study about general applications of digital human avatars. Our literature review, however, investigates the strategies toward building VCP environments. We also carried out a significantly more comprehensive reviewing survey than \cite{hudson2016avatar} by categorizing different methods, investigating their advantages, limitations and applications, comparing various approaches and categories and evaluating the evolution of this new technology.  

\FloatBarrier

This research's \textbf{contributions} are: This literature review is the first research that investigates different approaches toward building VCP environments upon digital human avatar models to the best of our knowledge. Notably, the categorization methodology of avatar-based methods in this paper is new. 
Moreover, we conducted a comprehensive analysis on different categories and methods based on their advantages, limitations, applications, and the evolution of technologies used in VCP environments. 

\section{Non-Avatar-based Approaches}
\label{Sec:non-avatar}
Non-avatar-based approaches usually rely on physical hardware to create a VCP environment. The hardware can be an interactive robot, a mobile system installed on a moving platform, or traditional video-based telepresence. While this research's primary focus is on avatar-based approaches, we briefly explain non-avatar methods as follows:       

\subsection{Robotics}
Robotics has been used for various VCP/telepresence purposes, such as communication, environmental visualization, and training physicians.  
However, its widespread usage has been to improve the interactions of a social VCP environment. 
For example, in \cite{minato2012development}  Minato et al. suggested a portable human-like to users can feel others' presence. This human-like sensation is achieved by using human-like voice, appearance, and touch that users can communicate and talk to others remotely. The experiments showed that people quickly started conversation with the robot and are impressed by its shape and feeling. 
In \cite{yoon2015control}, Yoon et al. proposed a robotic telepresence system with some modern features such as a projector and a head tracker system. These features make the communication between the user and robot more interactive. The suggested features are unique in telepresence applications that led the system to be more effective than traditional robotics system. 
Robotics also has been used for environmental visualization. For instance, in \cite{macharet2012collaborative} Macharet et al. designed a telepresence robot to visualize the environment for users. The robot is controlled by a remote human operator and can smoothly navigate through a house with capability of handling complicated situations such as narrow corridors and doors. The results showed that using the proposed method helps to reduce the number of  environmental collisions during navigating with the robot.

The main drawbacks of using robots are the need for regular maintenance, limitation of the robot to perform human-like interactions, the difficulty of adding new features and users' discomfort to use robots.

\subsection{Mobile Systems}
A mobile system is a set of interactive tools integrated on moving frames to create a VCP environment. Compared to conventional robots, mobile systems are more focused on mobility, accessibility, and practical applications \cite{parker2016SHR}. 
As an example, in ~\cite{beer2011mobile} Beer et al. designed a mobile system consisting of several modules such as a touch screen installed on a phone frame, a microphone, a web camera, and speakers to help elders to interact with visitors. The experimental results showed that elders have good experience of interactivity and visibility in the designed mobile system that will reduce traveling costs and social isolation. 
In ~\cite{lee2011now} Lee et al. suggested a mobile system to enable remote workers to live and work with local coworkers similar to how they do it physically in real life. They exploited a Texai Alpha prototype \cite{texai} mobile system for this purpose. The experimental results obtained from the surveys indicated that the remote pilots have a similar experience of working with real local coworkers. 

While mobile systems can be equipped with state-of-the-art electronic tools, they share the similar problems with robotics systems.  Moreover, they strictly rely on manufacturer's available development kit tools for adding any new features \cite{hening2013ICPTAE}.   

\subsection{Images and Videos}

Using images and videos is the traditional way of creating a co-presence environment. As an example of such an approach, in \cite{noda2015implementation} Noda et al. proposed a telecommunication system to connect different users using a configurable tile display. The proposed scheme includes several features to offer a realistic sensation of co-presence such as life-size processing, subtracting the background, and using multiple cameras. The designed VCP system provides a higher sense of presence compared to the traditional video-based approaches. 
An advanced video-based VCP environment is designed in \cite{maeda2004real} called the “real-world” where users can view each other from different perspectives. The real-world implementation is accomplished by using a multiview video capturing system and eighteen PCs in a cylindrical chamber. The experimental results showed that the suggested system can be delivered efficiently to users who watch others in real-time. 
Compared to robotics and mobile systems, using images and videos for creating VCP environments has some advantages and drawbacks. For example, while large displays can provide more human-like sensation and require minimum maintenance, they lack interactivity and a real 3D experience.  

Table \ref{Tab:nonavatar} summarizes the different non-avatar-based methods have been suggested to create VCP environments.

\begin{figure*}[h!tbp]
	\centering
	\begin{tabular}{c c c}
	\includegraphics[height=0.23\textwidth]{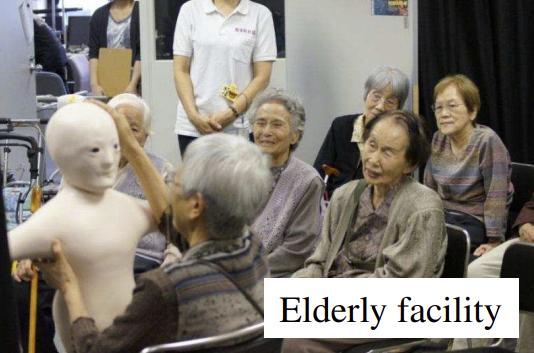} &
	\includegraphics[height=0.23\textwidth]{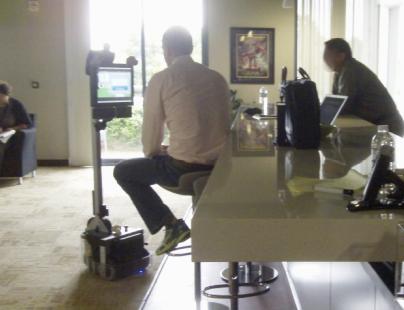} &
	\includegraphics[height=0.23\textwidth]{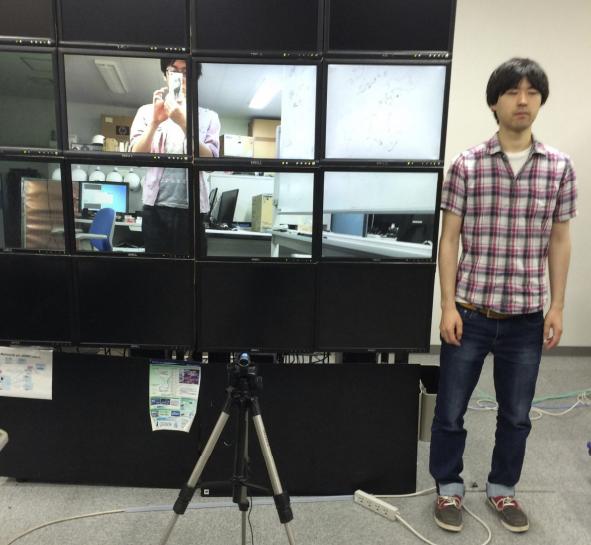} \\
	(a)  & (b) & (c) \\
	\end{tabular}
	\caption{Examples of (a) robotics \cite{minato2012development} , (b) mobile system ~\cite{lee2011now} and (c) image and videos-based \cite{noda2015implementation} VCP environments. 
	\label{Fig:non-avatar}} 	
\end{figure*}

\begin{table}[h!tbp]
	\centering
	\caption{Summary of non-avatar methods.
	~\label{Tab:nonavatar}} 
	
\begin{tabularx}{\linewidth}{LLLL} 
\hline
Category &  Applications &  Advantages & Disadvantages\\ \hline
Robotics  &  telepresence \cite{minato2012development,yoon2015control} and remote navigation~\cite{macharet2012collaborative} 
& compactness and configurable modular systems & regular maintenance, limited interactions and features and users' discomfort \\ \hline
Mobile Systems &  elderly healthcare assistance~\cite{beer2011mobile} and remote working~\cite{lee2011now} & mobility, accessibility and practical applications & regular maintenance and limited features and interactions \\ \hline
Images and Videos & tele-communication \cite{noda2015implementation,maeda2004real} & large displays and minimum maintenance & lack of interactivity and real 3D experience
\\ \hline
\end{tabularx}
\end{table}

\section{Avatar-based Approaches}
\label{Sec:avatar}
In avatar-based approaches, commonly, a digital 3D human model represents real humans in VCP environments \cite{hasler2013CHB}. Compared to non-avatar methods, using an avatar doesn't require any special maintenance and offers more human-like sensations and interactions in a 3D world. 
Moreover, in contrast to non-avatar approaches that rely on hardware, avatar-based strategies depend on software to create a VCP environment. Hence, they are more flexible in adding new features and can be upgraded based on state-of-the-art AI and Computer Graphics algorithms to fulfill users' needs.
In this paper, we first categorize the avatar-based approaches to two types of \emph{Pre-rendered motions} where the animations are pre-defined and \emph{Direct motion retargeting} where users directly control the avatars. We will give more explanation about the aforementioned categories as follows: 

\subsection{Pre-Rendered Motions}
\label{Sec:pre-render}
A 3D avatar in a VCP environment can be animated by pre-designed motions and scenes. For example, in \cite{pazour2018virtual}, Pazour et al. simulate a conference room with user-defined avatars that can communicate with each other remotely. This simulation's primary goal is to evaluate users' realistic feeling of co-presence in a virtual environment. Two experiments were conducted, where two scenarios control head motions: (1) tracking by a Head-Mounted Device (HMD), HTC Vive, and (2) using mouse movements in a desktop PC. The avatars' upper and lower body are animated using pre-rendered animations developed by Mixamo~\cite{Mixamo}. The experimental results showed that using the VR headset outperforms the desktop (mouse movement) in terms of feeling other users' presence.

Pre-rendered motions have also been used to simulate metropolitan structures such as college \cite{monahan2008virtual}, airport \cite{li2016virtual} and Museum \cite{mu2009implementation}. For example, in \cite{monahan2008virtual}, Monahan et al. simulated a college environment with students and teachers. In the designed scenario, students select a unique 3D character during the registration process to represent them onscreen in the 3D university environment. The avatars are human-like and can perform a variety of pre-designed actions associated with their roles. The results illustrated the effectiveness of the proposed technique, and users' realistic feeling of others' activities and presence.
In another example, in \cite{li2016virtual}, Li et al. simulated taking off an aircraft in an airport runway environment. Five users can interact with each other while wearing VR goggles to see other users` avatars. The users can take the role of a pilot, a support office, a tractor guide, a carrier aircraft guide, and a tractor driver. The carried surveys illustrated that the proposed type of exhibit is exciting and attractive for users. 
In \cite{mu2009implementation} Mu et al. designed a VR environment for multi-user learning in Museum. The primary way of the interaction between users is pre-defined gestures that can be customized by users using a Graphical User Interface (GUI). Participants can select their appearance and clothing before entering the VR environment. The results indicated that the method was effective for users to transfer their motion in the virtual location.
 
Pre-rendered motions also have been used for psychological testing purposes. For example in ~\cite{brown2017coordinating}, Brown et al. presented a narrative story in a VR environment. Users can play the game over a network connection wearing a head-mounted display (Oculus Rift). The goal is to study a set of guided camera and gaze distracting techniques to determine how to attract unfocused individuals to the same story. The results showed a better understanding of factors that cause users' attraction to a narrative story.
 
Using pre-rendered motions and scenes can be suitable for the scenarios that specific animations satisfy users' needs. However, in most cases, a more authentic approach to create movements directly from users' actions is desirable. To effectuate this, many researchers directly retarget motions from users to 3D avatars.

\subsection{Direct-Motion Retrageting}
In this category, the motion data is directly transferred from humans to 3D avatars. The surrounding environment can be pre-designed \cite{shapiro2014automatic,li20193d,beck2013immersive} or reconstructed scenes \cite{newcombe2015dynamicfusion,orts2016holoportation}. We divide the Direct-motion retargeting strategies into two types of \emph{Image-based} methods, where the inputs are image, video, or depth data and \emph{Sensors-based} approaches where the inputs are obtained from sensor equipment. 
While some of the researches in this category are directly implemented in VCP environments, some others suggested avatar modeling pipelines that are profoundly beneficial toward creating VCP environments. 

\begin{figure*}[h!tbp]
	\centering
	\begin{tabular}{c c}
	\includegraphics[height=0.28\textwidth]{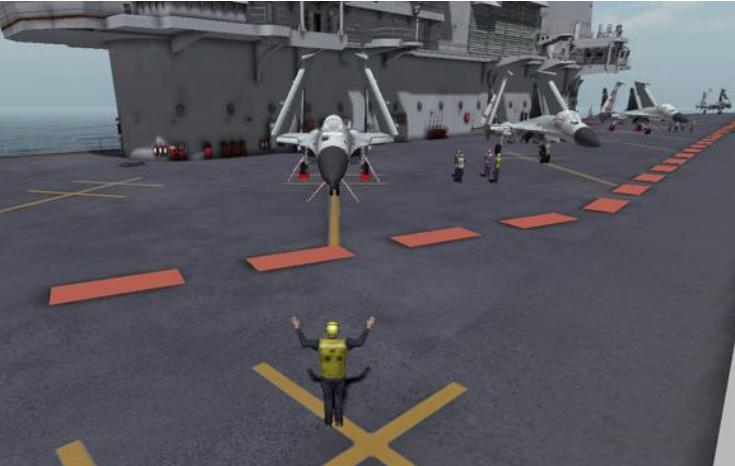} &
	\includegraphics[height=0.28\textwidth]{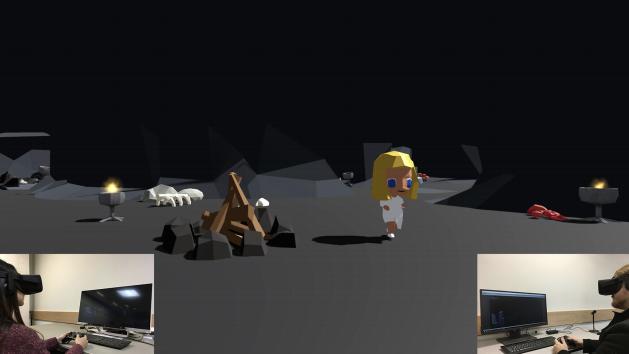} \\
	(a)  & (b) \\
	\end{tabular}
	\caption{Pre-rendered motions used for creating VCP environments to simulate an airport~\cite{li2016virtual} and a psychological test~\cite{brown2017coordinating}.
	\label{Fig:pre-rendered}} 	
\end{figure*}

\textbf{Sensor-based Approaches}

\label{Sec:sensor}
To animate human avatars, different sensors have been used; such as wearable sensors \cite{lifkooee2019real,han2017simulating,andreadis2010real}, head-mounted sensors \cite{herder2019avatars,wang2019effect,rauter2019augmenting}, and motion capturing systems with tracking markers~\cite{camporesi2015effects}. 

There have been different applications for sensor-based human avatar animation generation. Some researchers utilized sensors to evaluate the effectiveness of human avatars \cite{wang2019effect,camporesi2015effects,han2017simulating}. As an example, in \cite{wang2019effect} Wang et al. adopted HMD sensors (HTC Vive) to investigate the performance of different segment levels of a human avatar such as partial hands, full hands and full-body avatar. The experimental results indicated that the full-body avatar leads to the highest performance and satisfaction when they used HMD sensors. 
As another example, in \cite{camporesi2015effects} Camporesi et al. evaluated the performance of avatars and non-avatars techniques using a motion capturing system with ten cameras that can track users' head and body based on trackable markers. The results indicated that exploiting avatar-based categories can improve users' quality and speed on finishing their assigned tasks. 
In \cite{han2017simulating} Han et al. stimulated human's upper body motions using head and hand motion capturing sensors whose data are transferred in two different channels. The experimental results show an 80\% consistency rate between real human actions and digital avatar motions. 
In \cite{herder2019avatars}, Herder et al. created a VCP environment to simulate a large factory with industrial machines. Users' roles are as new workers and are trained based on a basic tutorial. Participants are tracked and animated in real-time using HMD sensors and development kits provided by HMD manufacturers. The results indicated that using a human avatar helps stimulating communication between users by having high immersive interactions and engagements. 

Sensors-based animated avatars also have been used to simulate social events \cite{andreadis2010real,de2019watching}. 
For example, in \cite{andreadis2010real} Andreadis et al. suggested a VCP environment to simulate a theater containing actors and other scenery subjects that are streamed in a multi-screen display. While the main actors' interactions are captured using a real-time motion capturing system and wearable magnet sensors, other subjects such as animals are animated using automatic AI-assisted motions. The effectiveness of the proposed VCP environment is validated by receiving positive feedback from both regular audiences and experts.
As another example, in \cite{de2019watching}. De et al. utilized Facebook Space, a pre-designed commercial software, to evaluate user's interaction. Participants interact with other human-like avatars by talking to them and listening to their conversation while watching movies. The final results showed that the users have similar experiences of watching movies together when using 3D avatars and traditional video-based environment.

Another application of sensor-based animated avatars is to analyze human behavior and motions in different scenarios~\cite{park2019investigation,rauter2019augmenting}.
As an example, in \cite{park2019investigation} Park et al. created a system to capture and analyze users' walking-in-place movement in VR environments. The motion sensors are installed on users' lower legs, and the avatar's lower leg is animated when any motion is detected. They concluded that the avatar's movement is natural and accurate. 
In \cite{rauter2019augmenting}, a mixed-reality environment is designed to investigate the interaction between users and objects. They combined the real world and virtual objects using depth sensors and an HTC Vive Pro HMD. The experimental results indicated that creating such a scenario to interact with virtual near-real objects in a mixed-reality environment is highly viable for real-world applications. As another instance, in \cite{maloney2019dc} Maloney suggested a VR environment where users can embody the avatar with different races to measure their racial bias. Users need to shoot human-like targets that are explained to be aliens invading the earth. They concluded that the proposed strategy is successful in decreasing users' implicit bias against different races. 

Compared to pre-rendered animations, generating motions by using sensors is more flexible, allowing users to perform various actions. However, using sensors has some limitations such as the need of proper maintenance, calibration, training and difficulty of wearing or using the sensor devices.   

Both of senor-based and pre-rendered motion-based approaches are more focused on application rather than techniques. So, in Table \ref{Tab:avatar} we compare these two categories based on their applications, advantages and disadvantages.  

\begin{figure*}[h!tbp]
	\centering
	\begin{tabular}{c c}
	\includegraphics[height=0.32\textwidth]{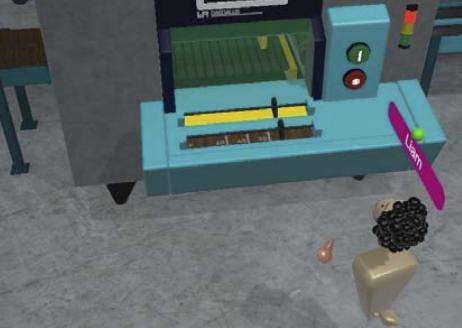} &
	\includegraphics[height=0.32\textwidth]{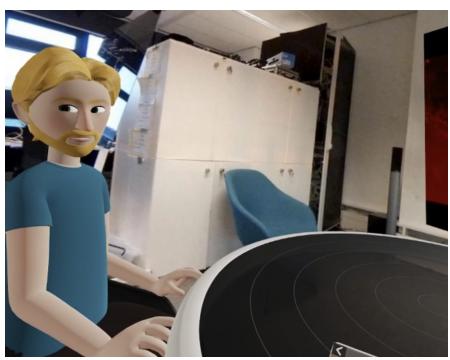} \\
	(a)  & (b) \\
	\end{tabular}
	\caption{Sensor-based VCP environments created to (a) simulate a large factory~\cite{herder2019avatars} and (b) watching movies~\cite{de2019watching}.
	\label{Fig:sensors}} 	
\end{figure*}

\begin{table}[h!tbp]
	\centering
	\caption{Comparison between Pre-rendered motion and Sensor-based (direct motion retargeting)   
	~\label{Tab:avatar}} 
	
\begin{tabularx}{\linewidth}{LLLL} 
\hline
Category &  Applications &  Advantages & Disadvantages\\ \hline
Pre-rendered motions  & virtual conference room   \cite{pazour2018virtual}, simulating metropolitan structures: university~\cite{monahan2008virtual}, airport~\cite{li2016virtual} and Museum~\cite{mu2009implementation}, psychological test \cite{brown2017coordinating} & minimal motion transfer error, ease of design based on the application, no need for maintenance and calibrations or wearing extra equipment & limited number of interactions and lack of authentic motion directly transferred from users\\ \hline
Sensor-based direct motion retargeting  &  evaluating the effectiveness of human avatars~\cite{wang2019effect,camporesi2015effects,han2017simulating}, simulating social events~\cite{andreadis2010real,de2019watching}, analyzing human behavior and motions~\cite{park2019investigation,rauter2019augmenting} & authentic motion transfer directly from users, no limitation on variety of motions  & need for regular maintenance, calibration and training and difficulty of using wearable devices\\ \hline
\end{tabularx}
\end{table}

\textbf{Image-based Approaches}
\label{Sec:image}

Images (including RGB and depth) have been widely used as the primary inputs for motion synthesis and retargeting \cite{wang2019generative,yang2020transmomo,lifkooee2018image,doersch2019sim2real}. As a result, images can be utilized as the inputs to animate human avatars in a VCP environment. 
In contrast to wearing/using sensors, capturing images does not require specific maintenance, training, and calibration. Moreover, users have more freedom to perform any action without the limitations caused by wearable devices.  
Image-based strategies can be further divided into two Offline and Online retargeting approaches. In Offline methods, the process of retargeting can be computationally expensive and might not be suitable for real-time applications. On the other hand, they might offer better accuracy by reconstructing human's body shape, pose, face, and textures. Nevertheless, Online motion retargeting is capable of being used in real-time since they are optimized and sometimes implemented with minimal realization~\cite{orts2016holoportation}. 
We summarize the comparison between different image-based strategies in Table \ref{Tab:image} and then will explain each sub-category with details as follows.

\begin{table}[h!tbp]
	\centering
	\caption{Brief comparison of different Image-based Direct motion retargeting approaches  
	~\label{Tab:image}} 
	
\begin{tabularx}{\linewidth}{LLLL} 
\hline
Category &  Applications & Advantages & Disadvantages\\ \hline
Image-based Offline motion retargeting & non-real time applications such as creating pre-rendered avatars, repairing avatar model damages caused by scanning error &  accurate and high resolution geometry reconstruction and texturing, accurate offline inpainting to fix the 3D model damages & expensive computational cost and difficulty of being implemented in real-time applications   \\ \hline
Image-based Online motion retargeting (Rigged avatar)  & VCP environments with generic avatars such as workers in training environments, system with simple setup & low computational cost, simple system setup, flexibility of selecting the avatars based on the application  & lack of authentic avatar that resembles the user's appearance, motion retargeting error caused by different skeleton structures \\ \hline
Image-based Online motion retargeting (3D model reconstruction) & VCP environments with avatar resembling users: tele-presence, tele-conference and meeting  & authentic avatar resembling users appearance, capability of reconstructing the surrounding enviroment in real-time  & high computational cost, complex system design and setup\\ \hline
\end{tabularx}
\end{table}

\textbf{Offline Motion Retargeting}
\label{Sec:offline}

Offline motion retargeting is mainly used to reconstruct high-quality human body shape, pose and textures \cite{bogo2015detailed,cui2012kinectavatar,korban20193d,lim2014rapid,malleson2017rapid,shapiro2014rapid,zhang2014quality}.  
As an example, in \cite{lim2014rapid} Lim et al. fused several approaches such as KinectFusion \cite{pagliari2014kinect} and ICP-based registration algorithm to generate the 3D avatar. The proposed method can create a human avatar with textures automatically. They enforce positional constraints to avoid motion artifacts. However, the ICP-based registration algorithm might not be suitable for highly deformable subjects such as humans. 

To overcome such a non-rigid deformation in human body geometry,  \cite{cui2012kinectavatar}, Cui et al. suggested an approach based on a non-rigid registration algorithm to reconstruct 3D human body geometry. First, several images are captured using a depth camera (Microsoft Kinect). Then, multiple stages mainly including a super-resolution-based algorithm and the Poisson mesh reconstruction \cite{kazhdan2006poisson}, are suggested to reconstruct the human body geometry and textures. The proposed method can automatically reconstruct non-rigid subjects such as humans smoothly and with high accuracy due to adding color constraints. Nevertheless, the suggested approach might be sensitive to issues such as holes or missing regions and soft tissue deformation since they utilized low-dimensional statistical models.  

To handle the soft tissue deformation and hole filling problems, in \cite{bogo2015detailed} Bogo et al. exploited both high and low dimensional models to reconstruct human body shape and pose from RGB-D video sequences. While the low dimensional model is used for initial pose estimation, the high dimensional model is utilized to repose and accurately reconstruct the geometry. The reconstruction process is done by displacement mapping between the local and global geometries. The results showed that their method is reliable even in the challenging situations such as varying resolution and soft tissue deformation. Still, the facial details can not be encoded accurately because the statistical model is built upon human body landmarks. 

There have been some methods that developed strategies to encode human facial features to be used in VR environments. These methods have been based on recognizing Action units \cite{vicente2019development,lifkooee2019video}, motion tracking of facial landmarks \cite{kegel2020dynamic} and bone marker tools \cite{el2019open}. Yet, these approaches only render faces while most VCP environments require full human body avatar models. 

To include effective facial features in full human body avatar models, in \cite{malleson2017rapid} Malleson et al. designed a system to create full-body avatars replicating the person's body shape, face and textures. When the body shape is created using blendshapes based on the body dimensions obtained from the depth images, the face is reconstructed using blendweights and the facial landmarks obtained from images. The experimental results illustrated that the reconstructed avatars look real enough for users to feel other's real presence in a VR environment. However, the computational cost of the whole reconstruction process is expensive (around 10 seconds).
We summarize the evolution of technology (as mentioned above) in the Image-based Offline motion retargeting category in Table \ref{Tab:offline}. 

\begin{table}[h!tbp]
	\centering
	\caption{Gradual evolution of technology in the Image-based Offline motion retargeting approaches  
	~\label{Tab:offline}} 
	
\begin{tabularx}{\linewidth}{LLLL} 
\hline
Paper/Year &  Solved issue(s) from previous paper(s) & Proposed method to solve the issue(s) & Other used algorithms\\ \hline
 Cui et al.~\cite{cui2012kinectavatar} 2012 & highly deformable human body & 
non-rigid registration algorithm & super-resolution-based algorithm, Poisson -based mesh reconstruction \\ \hline

Bogo et al.~\cite{bogo2015detailed} 2015 & soft tissue deformation ~\cite{cui2012kinectavatar} & high dimensional models & low dimensional model for initial pose estimation, displacement mapping \\ \hline

Malleson et al.~\cite{malleson2017rapid} 2017 & lack of encoding facial landmarks~~\cite{bogo2015detailed} & blendweights and facial landmarks & blendshapes for body shape reconstruction \\ \hline
\end{tabularx}
\end{table}

\textbf{Online Motion Retargeting}
\label{Sec:online}

Compared to Offline motion retargeting, Online methods are optimized and can be utilized in real-time. We categorize the Online motion retargeting strategies into two types of Rigged-avatar-based motion retargeting and 3D avatar model reconstruction. We will give more explanation as follows:

\textbf{Rigged-Avatar-Based Motion Retargeting}:
In the Rigged-avatar-based motion retargeting category, the main focus is on the quality of motion transfer rather than quality of 3D avatar models. In this group, often pre-designed 3D avatar models depending on the application are used. As a result, they are more suitable for VCP environments that don't require the human avatars resembling users' appearance.  They also can be implemented cheaper and faster. 

For example, in \cite{jo2014avatar} Jo et al. created a 3D teleconference in an Augmented Reality (AR) environment using a generic rigged avatar model whose motion information is obtained by the Microsoft Kinect. They preserve the spatial property of objects that users can sit on digital chairs similarly to a real environment. The obtained surveys indicated that the users have more impressive and realistic experience than the traditional video-based communication approaches. Still, they evaluated their AR system based on limited motions, while complex movements are inevitable in a real-time VCP environment. 

This limitation \cite{jo2014avatar}, is solved in \cite{lugrin2015avatar} and \cite{choi2019effects} by considering various motions to have a more reliable interactive environment. Specifically, in, \cite{lugrin2015avatar} Lugrin et al. suggested a strategy to evaluate the impact of different types of avatars on the performance of fitness training. They exploited the Microsoft Kinect to capture the motion data in real-time and transfer it to human-like rigged avatars. 
As another example, in \cite{choi2019effects} Choi et al. evaluated the impact of different types of motions on users’ feelings of body ownership and presence. They utilized a OptiTrack Motion Capturing System \cite{opt} with six cameras for real-time motion tracking and retargeting. 
The experimental results indicated the effectiveness of these approaches \cite{lugrin2015avatar,choi2019effects} for training and testing purposes, respectively. However, the Kinect or motion capturing system cannot extract the facial features effectively because of the limited resolution of commercial motion cameras. 

To tackle the shortage of facial features \cite{lugrin2015avatar,choi2019effects}, in \cite{roth2017socially} Roth et al.  designed an immersive environment that uses a tracking system (OptiTrack sensors \cite{opt}), an eye gaze tracking and facial expression tracking (Faceshift software) to evaluate users' social interactions. The proposed strategy offers various interactive features that help users' immersion in the environment based on a stereoscopic projected display. Still, the real-time tracking errors caused by complex motions are not evaluated while these errors can negatively affect a high quality VCP experiment. 

To reduce the artifacts caused by complex motions such as turning around or partial view Kang et al. \cite{kang2019real} suggested an adjustable filter integrated with multiple Inverse Kinematics (IK) constraints. The motion information is obtained from a Kinect device and is transferred to a rigged avatar using quaternions calculations. 
We summarize the aforementioned evolution of technology in the Rigged-avatar-based Online motion retargeting group in Table \ref{Tab:rigged}. 

\begin{figure*}[h!tbp]
	\centering
	\begin{tabular}{c c}
	\includegraphics[height=0.5\textwidth]{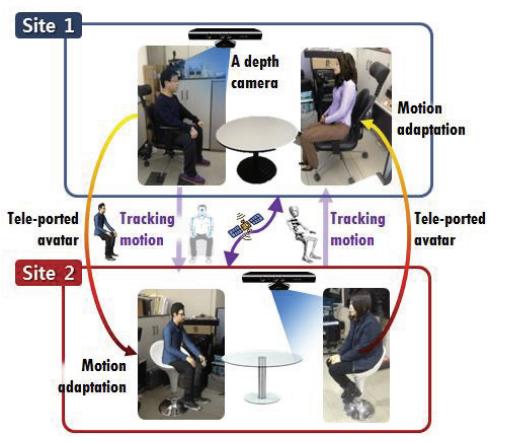} \\
	\end{tabular}
	\caption{Rigged avatars used to simulate a 3D tele-conference \cite{jo2014avatar} 
	\label{Fig:rigged}} 	
\end{figure*}

\begin{table}[h!tbp]
	\centering
	\caption{Gradual evolution of technology in the Rigged-avatar-based Online motion retargeting approaches   
	~\label{Tab:rigged}} 
	
\begin{tabularx}{\linewidth}{LLLL} 
\hline
Paper/Year &  Solved issue(s) from previous paper & Proposed method to solve the issue(s) & Application \\ \hline
Jo et al.~\cite{jo2014avatar} 2014 & preserving spatial properties & global and local motion adaptation & tele-conference \\ \hline
 Lurgin et al. ~\cite{lugrin2015avatar}  2015 & limited motions ~\cite{jo2014avatar} & 
 advanced tracking  & virtual fitness training \\ \hline
 Roth et all. ~\cite{roth2017socially} 2017 & lack of  facial features  ~\cite{lugrin2015avatar} & Faceshift software & virtual social gathering \\ \hline
Kang et al. ~\cite{kang2019real} 2019 & artifacts in complex motions~~\cite{roth2017socially}  & adjustable filters & virtual fitness training\\ \hline
\end{tabularx}
\end{table}

\textbf{3D Model Reconstruction:}
3D reconstruction of avatar models has some advantages over rigged avatars, such as generating avatars that resemble users' body shape, face, and reconstructing the environmental scene. In this category, some researchers suggested approaches to reconstruct the human 3D model \cite{shapiro2014automatic,li20193d,beck2013immersive} or the whole scene \cite{newcombe2015dynamicfusion,orts2016holoportation}. The image-based 3D Model reconstruction category often includes the most advanced state-of-the-art algorithms for human avatar modelling. 

As an example in, \cite{shapiro2014automatic} Shapiro et al. combined several methods such as KinectFusion \cite{newcombe2011kinectfusion} and Poisson surface reconstruction algorithm \cite{kazhdan2006poisson} to reconstruct the human body geometry and animation based on key-poses and superresolution range scana. Although they can reconstruct the whole human geometry in a few minutes, the scanning process and auto-rigging of the reconstructed mesh are performed offline. As a result, the entire motion retargeting process can not be implemented in real-time.

To resolve the offline scanning issue \cite{shapiro2014automatic}, in \cite{newcombe2015dynamicfusion}, Newcombe et al. proposed a scene reconstruction algorithm that can reconstruct the whole scene in real-time. The suggested approach is based on a DynamicFusion scheme consisting of three stages of parameter estimation, fusion and structure adaptation, to fuse the RGBD data and reconstruct the subjects. The suggested strategy can handle reconstructing highly deforming subjects such as human body in real-time. Nevertheless, they did not suggest any solution to add textures and handle the tracking errors caused by complex geometry, partial view, occluded or shadowed parts. 

To resolve the tracking errors and real-time texture mapping issues, \cite{shapiro2014automatic,newcombe2015dynamicfusion}, Orts et al. \cite{orts2016holoportation} developed a system called Holoportation that can handle the occlusion using a temporal reconstruction algorithm. The proposed system includes multiple RGB and infrared cameras to capture and transmit the moving human body's dynamic 3D geometry and the surrounding scene. While the method can reconstruct high-quality human body 3D models with textures, there are some drawbacks such as expensive and sophisticated setup due using multiple RGB and depth cameras to reconstruct the whole scene. 

To overcome the difficulty of complex setup \cite{shapiro2014automatic}, in \cite{li20193d} Li et al. proposed a strategy to reconstructs human body model and textures from a single RGB image. The suggested approach includes multiple stages, including human body segmentation, fitting the segmented body to a parametric model,  wrapping the initial geometry to the final model based on a dense correspondence and silhouette. They suggested a new network called InferGAN to interpret the textures of invisible parts from users behind. Yet, they concluded some limitations such as limited camera view and sensitivity to occluded body limbs.  We summarize the aforementioned evolution of technology in the 3D model reconstruction-based online motion retargeting category in Table \ref{Tab:model}  

\begin{table}[h!tbp]
	\centering
	\caption{Gradual evolution of technology in the 3D model reconstruction-based online motion retargeting approaches   
	~\label{Tab:model}} 
	
\begin{tabularx}{\linewidth}{LLLLL|} 
\hline
Paper/Year &  Solved issue(s) from previous paper & Proposed method to solve the issue(s) & Other used algorithms \\ \hline
 Shapiro et al. ~\cite{shapiro2014automatic} 2014 & slow human geometry reconstruction ~\cite{lim2014rapid} & representative frames (key-poses) & Poisson Surface Reconstruction, superresolution range scanning algorithm \\ \hline
 Newcombe et al.~\cite{newcombe2015dynamicfusion} 2015 & offline scanning ~\cite{shapiro2014automatic} & DynamicFusion (online-scan-streaming) & parameters estimation, fusion and structure adaptation \\ \hline
 Orts et al. ~\cite{orts2016holoportation} 2016 & real-time tracking errors and texture mapping ~\cite{newcombe2015dynamicfusion,shapiro2014automatic} & temporal reconstruction algorithm and textures compressing & dynamic 3D geometry reconstruction \\ \hline
 Li et al ~\cite{li20193d} 2019 & expensive and complex setup ~\cite{orts2016holoportation}  & single RGB image-based human body reconstruction and texturing & InferGAN to interpret the textures of invisible parts, parametric model \\ \hline
\end{tabularx}
\end{table}

\begin{figure*}[h!tbp]
	\centering
	\begin{tabular}{c c c}
	\includegraphics[height=0.23\textwidth]{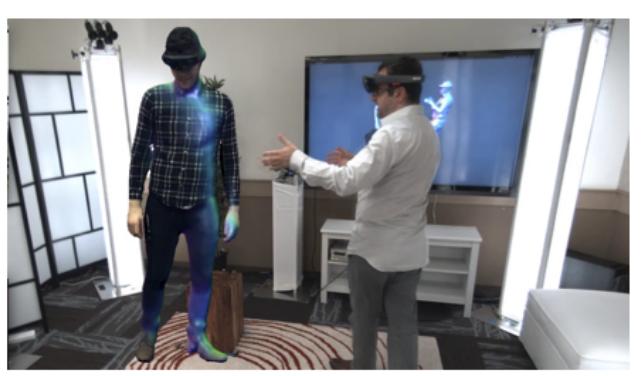} &
	\includegraphics[height=0.23\textwidth]{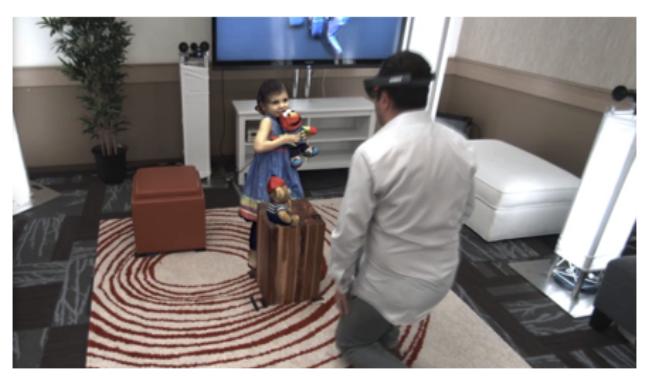} \\
	\end{tabular}
	\caption{Online Image-based 3D reconstruction of avatar and scene models ~\cite{orts2016holoportation}, 
	\label{Fig:holo}} 	
\end{figure*}

\FloatBarrier

\section{Conclusion}

In this survey paper, we reviewed the methods which created and used human avatar models toward designing VCP environments. After a short discussion about the non-avatar strategies, we discussed the avatar-based techniques, their advantages and disadvantages, and the gradual advancement of technology for each method and category. 
We conclude that the recent advancement in computer graphics/vision algorithms profoundly has improved the quality of human avatar representation in VCP environments.

\bibliographystyle{unsrt}

\bibliography{main.bib}

\end{document}